# Sub-micron spatial resolution in far-field Raman imaging via positivity constrained super-resolution

**Dominik J. Winterauer[1,2], Daniel Funes-Hernando[2], Jean-Luc Duvail[2], Saïd Moussaoui[3], Tim Batten[1] and Bernard Humbert[2]**

## Abstract

Raman microscopy is a valuable tool for detecting physical and chemical properties of a sample material. When probing nanomaterials or nanocomposites the spatial resolution of Raman microscopy is not always adequate as it is limited by the optical diffraction limit. Numerical post-processing with super-resolution algorithms provides a means to enhance resolution and can be straightforwardly applied. The aim of this work is to present interior-point least squares (IPLS) as a powerful tool for super-resolution in Raman imaging through constrained optimisation. IPLS's potential for super-resolution is illustrated on numerically generated test images. Its resolving power is demonstrated on Raman spectroscopic data of a polymer nanowire sample. Comparison to AFM data of the same sample substantiates that the presented method is a promising technique for analysing nanomaterial samples.

[1]Renishaw plc
[2]Institut des Matériaux Jean Rouxel Nantes (IMN)
[3]Laboratoire des Sciences du Numérique de Nantes

**Corresponding author:**
Dominik J. Winterauer, Renishaw plc, New Mills, Wotton-under-Edge, Gloucestershire, GL12 8JR, United Kingdom. Tel: +44 1453 524524
Email: dominik.winterauer@protonmail.com

**Introduction**

Super-resolution image reconstruction (SRIR) uses signal processing techniques to overcome the resolution limitation of conventional (optical) imaging apparatus[1], usually by combining multiple images of the same scene[2]. A multitude of numeric SRIR methods based on different philosophies are available in the literature[1] and various super-resolution algorithms have been employed for SRIR in Raman spectroscopy mapping over the last decade[3–6]. The application of SRIR to Raman spectroscopy is a promising method that allows distances and objects below the diffraction limit to be resolved using far-field instrumentation, without the added experimental complexity of near-field techniques such as nearfield Raman spectroscopy (NFRS)[7,8] or tip-enhanced Raman spectroscopy (TERS)[9,10]. While the concept of SRIR is already very powerful in itself, it still takes adequate optimisation algorithms to unleash its full potential.

A promising approach to SRIR is constrained optimisation, which has been successfully applied in image inpainting[11], non-negative matrix factorisation[12,13], and achieved great resolution enhancements in digital image restoration, whether the constraint was used explicitly[14], or implicitly by the choice of regularisation function[15–17].

In this work a primal-dual method for interior-point least squares (IPLS)[18] was used for constrained optimisation SRIR. IPLS was applied to Raman spectroscopic data acquired from a bundle of nanowires made of poly-(3,4 ethylenedioxythiophene) here referred to as PEDOT[19,20]. Atomic force microscopy (AFM) images were obtained from the same sample area. AFM imaging is not affected by the optical diffraction limit and hence provides a means to evaluate the quality of the super-resolved Raman images.

The discussion of the results focuses on the resolving power if SRIR, a point that was sometimes ambiguous in related work. The ambiguity arises from the two research disciplines overlapping in Raman SRIR, digital image restoration and microscopy, having slightly different notions of *resolution*. In digital image restoration resolution is usually seen as the level of detail contained in an image and often characterised by the sharpness of features in the image, a point of view that has been adopted in previous work on SRIR in Raman imaging[3–6]. Resolution in microscopy, on the other hand, is defined by the minimal separation between two objects necessary to observe them as distinct objects. SRIR results in Raman imaging have not been analysed from that perspective and while the two points of view are somewhat related they are not strictly identical.

The aim of this paper is to highlight the difference between these two interpretations of resolution and to demonstrate that Raman SRIR is capable of resolving distances smaller than the diffraction limit in the strict microscopist's sense.

**Materials and methods**

*Sample preparation*

PEDOT nanowires were synthesised by the hard template method described in a previous paper[20]. The template was an anodic aluminium oxide membrane of $(50 \pm 1)\mu m$ thickness and $(100 \pm 10)nm$ pore diameter (Synkera Unikera SM 100-50-13). The pore diameter constrains the diameter of the synthesised nanowires. The sonicated PEDOT nanowires were dispersed in water and drop cast on a silicon substrate[20]. The dispersion, as most dispersions of nano-objects in water, is incomplete and the nanowires tend to stick in bundles of different sizes, shapes and numbers of nanowires, although some of them were found isolated (see fig. 3a).

*Instrumentation*

Raman spectroscopic measurements were performed using a confocal Raman microscope (Renishaw inVia) with a $532nm$ frequency doubled Nd-YAG laser as the excitation source. A 100× 0.85 NA objective lens was used to focus the laser onto the sample and to collect the Raman back scattered light. The laser was linearly polarised along the vertical image axis. Laser power on the PEDOT sample was $0.1mW$ with acquisition time $0.5s$ per spectrum. With these parameters no bleaching or other photo induced modification of the sample was observed. The system was operated in high confocal mode to ensure the best lateral resolution was achieved. Raman mapping was conducted under the microscope using a motorised piezo-electric stage (Physikinstrumente Hera P621.2CD XY) in closed loop operation ($0.4nm$ resolution, $2nm$ repeatability). Data were collected by point mapping in a raster scanning fashion and each Raman spectrum is acquired under identical conditions. For this work a $20nm$ step size was used.

The Raman system with the piezo stage rests on an optical table that is passively damped and any environmental vibrations were not seen to be a significant issue. Measurement parameters were optimised to reduce measurement times to avoid artefacts originating from other environmental factors such as thermal drift.

Atomic force microscopy (AFM)[21] was performed with a scanning probe microscope (Bruker Innova) operated in tapping mode[22]. The AFM tips used were Olympus AC160TS-10 (OTESPA) AFM tips in visible apex geometry with a side and back angle of $35° ± 1°$ each, a front angle of $0° ± 1°$, and a maximum tip radius of $10nm$[23]. AFM imaging is a high-resolution imaging technique and served as validation of the results obtained by SRIR from the Raman data. For samples with high aspect ratios, and heights exceeding the tip curvature radius, the lateral resolution
of the AFM is adversely affected by the tip geometry[24–26]. Since the diameters of the studied PEDOT nanowires were well in excess of the tip radius, tip geometry effects were visible in the AFM images obtained. Tip geometry effects, however, only broaden the apparent width of objects, while position, height, orientation, and alignment are imaged faithfully and the obtained lateral resolution was still significantly better than that of the diffraction limited Raman microscope.

*Super-resolution image restoration*

SRIR relies on inverse problem resolution and thus needs three ingredients to work properly: an observation model, an optimisation criterion, and a powerful optimisation algorithm[27].

The observation model describes how an ideal high-resolution image transforms into a blurry and noisy image through the distortions imposed by the imaging apparatus and the imaging process. Observation models in
digital optical imaging can be written in the compact linear form[1,28,29]

$$y = Hx + \eta. \tag{1}$$

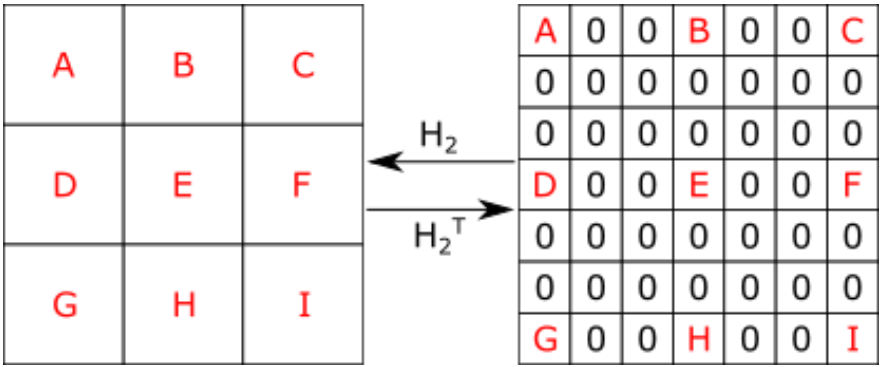

**Figure 1.** Illustration of the up- and downsampling process: The alphabetically labelled pixels in the low-resolution image (left) are identical to the alphabetically labelled pixels in the high-resolution image (right). The '0'-labelled pixels in the high-resolution image have no correspondence in the low-resolution image, these are the pixels that have been (artificially) inserted. They are equal to 0 when $H^T_2$ is applied to the low-resolution image.

Each index of column vector y corresponds to one of the $M$ pixels in the low-resolution image. Likewise, each index of the column vector x corresponds to one of the $N$ pixels in the high-resolution picture. The $M \times N$ matrix H is the *forward operator* which describes the weights with which each pixel intensity in the high-resolution picture x affects each pixel intensity in the low-resolution image y. The additive term $\eta$ accounts for any non-deterministic process during imaging and is usually modelled as independent and identically distributed (i.i.d.) Gaussian noise. For the present case the linear degradation H is factored into two terms[1]

$$H = H_2 H_1. \tag{2}$$

$H_1$ corresponds to a 2D convolution with the microscope's *point-spread function* (PSF) and $H_2$ represents downsampling. The former is ubiquitous in microscopy and gives rise to the well-known diffraction limit in conventional optical microscopes. The PSF depends on the specifications of the instrument and will be determined experimentally below.

If the high-resolution image x has more pixels than the observed lowresolution image y a decimation or downsampling operation $H_2$ has to be included in the observation model[1,29,30]. Conversely, its transpose $H^T_2$ is known as upsampling and describes the manner in which pixels are added to the low-resolution image. The upsampling operation $H^T_2$ is dependent on the pixel dimensions of the high-resolution image which in most cases can be chosen as desired. The corresponding downsampling operation $H_2$ is found by transposition. Additional pixels can be added with equal spacing along each dimension between existing pixels in the low resolution image y. The point mapping nature of the confocal imaging process means that there is a one to one correspondence between the lowresolution image pixels and a subset of the pixels in the high-resolution image. Fig. 1 shows the case where two high-resolution pixels are inserted between each two low-resolution pixels along each dimension.

SRIR in this paper is performed by interior-point least squares (IPLS). The optimisation criterion solved is

$$\hat{x} = \underset{x}{\mathrm{argmin}} (y - Hx)^T (y - Hx) \quad (3a)$$

$$s.t. \; \mathbf{x} \succeq 0, \quad (3b)$$

The positivity constraint on the pixel intensities, eq. (3b), reflects the natural assumption that abundances of scattering sources can only be nonnegative. Given y and H are known, eq. (3) can be solved by a primaldual optimisation method for IPLS[18]. SRIR via eq. (3) on numerically generated test data with known degradation H is demonstrated in the supplemental material.

*PSF measurement*

The instrument's PSF enters the forward operator H through $H_1$, eq. (2). Because H must be known for solving eq. (3), the PSF of the instrument has to be determined prior to any SRIR. The Raman microscope used in this work probes the sample with an excitation laser. It can be reasoned from the beam shape of lasers in resonators, the subsequent optics[31–34], and the response function of optical microscopes[35] that the intensity profile of the PSF is approximately Gaussian.

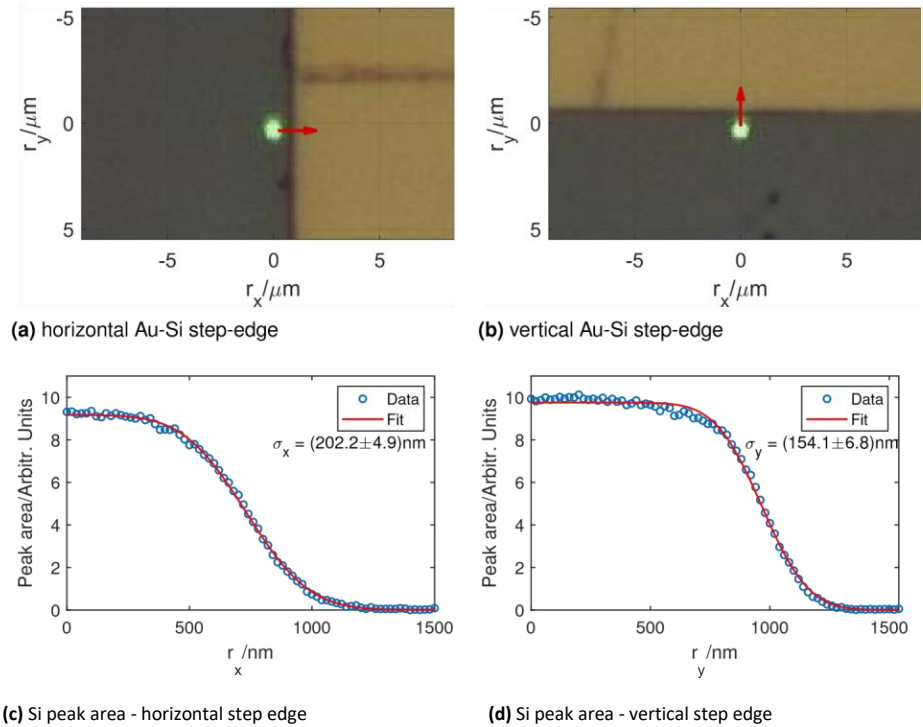

**Figure 2.** Measurement of the microscope PSF: Optical images of the horizontal (a) and vertical (b) Au-Si step-edge. The Raman signal is acquired along the red arrows with a sampling step of 20 nm. The area of the principle Si peak at $520.7 cm^{-1}$ is calculated for each sampling point and plotted against the distance (c, d). The model function eq. (5) is fitted to the data.

A general 2D Gaussian takes 3 independent parameters but from the symmetry of the instrument's configuration it can be assumed that its functional dependency factorises along the stage axes. Thus, the PSF is sufficiently described by

$$p(r_x, r_y) \propto \exp\left(-\left(\frac{r_x^2}{2\sigma_x^2} + \frac{r_y^2}{2\sigma_y^2}\right)\right), \quad (4)$$

where $r_x$ and $r_y$ are the coordinates along the stage axes. This is supported by the data shown in fig. 2. The parameters $\sigma_x$ and $\sigma_y$ in eq. (4) were determined experimentally using the step-edge criterion.

The laser was scanned over the edge of a gold pad deposited on a silicon substrate. (figs. 2a and 2b). Raman scattering only occurs when laser light falls on the silicon and not on the metallised region. The resulting intensity profile is a 1D line integral of eq. (4) that can be modelled by

$$\phi(r, a, b, \mu, \sigma) = a + \frac{b}{2}\left(1 + \mathrm{erf}\left(\frac{r-\mu}{\sqrt{2}\sigma}\right)\right). \quad (5)$$

For horizontal and vertical step edges σ in eq. (5) is equal to $\sigma_x$ and $\sigma_y$ in eq. (4), respectively. Fig. 2 shows the peak area profile of the principal silicon Raman peak across a horizontal and vertical step-edge. The peak area was calculated by integrating the Raman signal from $480 cm^{-1}$ to $562 cm^{-1}$ in the wavenumber domain, which contains the principal Raman peak of silicon (found at $520.7 cm^{-1}$). Prior to integration a linear background was subtracted from each spectrum to remove any interfering fluorescence signal. Eq. 5 is fit to each of the profiles and the values obtained for σ are

σ$_x$ = (202.2 ± 4.9) nm (6a) σ$_y$ = (154.1 ± 6.7) nm. (6b)

These correspond to full widths at half maximum (FWHM) of

FWHM$_x$ = (476.1 ± 11.5) nm　　　　　　(7a)

$$FWHM_y = (362.9 \pm 15.8)\ nm, \qquad (7b)$$

which can be used as a measure for the minimum resolvable distance of the instrument in the respective direction (see supplemental material for more details). The Raman data for PSF measurement were acquired with $20mW$ laser power at $532nm$, $0.2s$ acquisition time per spectrum and $20nm$ step size.

## Results

### Raman and AFM mapping of PEDOT nanowires

An optical image of the PEDOT nanowire sample area is shown in fig. 3a. A map of $1.2\mu m \times 1.5\mu m$ is collected from within the red rectangle displayed. The Raman data of PEDOT nanowires are acquired with $0.1mW$ laser power at $532nm$, $0.5s$ acquisition time per spectrum and a step size of $20nm$ along each stage axis. Fig. 3b shows a PEDOT Raman spectrum. The two main peaks at $1437cm^{-1}$ and $1518cm^{-1}$ are attributed to the symmetric and anti-symmetric vibration modes of the aromatic C=C bond, respectively[19].

The Raman intensity of the spectral channel centred at the peak position of the dominant C=C peak ($1437cm^{-1}$) is plotted as a heat map over the sampling area in fig. 4a. An AFM map of size $3\mu m \times 3\mu m$ covering the area probed by Raman spectroscopy was acquired for comparison. The number of sampling points was $512 \times 512$, resulting in a pixel size of approximately $6nm \times 6nm$. The measured height across the Raman probed area is shown in fig. 4b. Two nanowires separating from each other at the top of the mapping area, as visualised in fig. 4b, cause the spatial broadening of the Raman signal perpendicular to the direction of the wire observed in fig. 4a. However, the separation of the two wires within the sampling areas is smaller than any of the minimum resolvable distances of the Raman microscope, eq. (7), and hence they cannot be directly observed as individual wires.

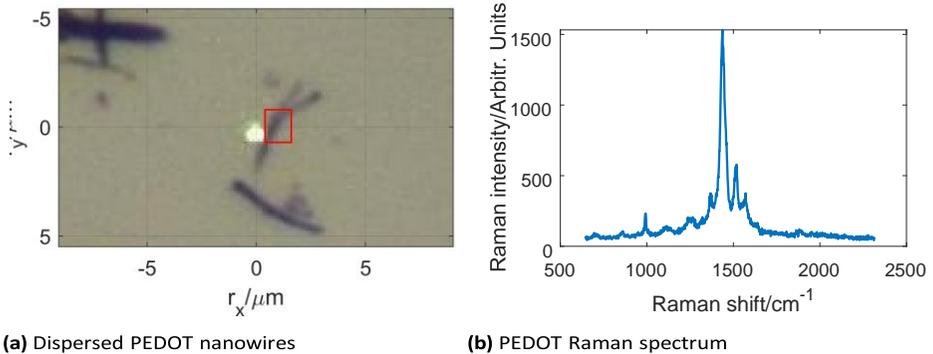

(a) Dispersed PEDOT nanowires  (b) PEDOT Raman spectrum

**Figure 3.** The PEDOT sample: (a) Optical image of PEDOT nanowires dispersed on a Si substrate: The Raman data is collected from within the red rectangle of $1.2\mu m \times 1.5\mu m$. The bright green dot at the origin is a visualisation of the $532nm$ laser spot. (b) PEDOT Raman spectrum: The dominant peak at $1437cm^{-1}$ is caused by the symmetric vibration mode of the aromatic C=C bond[19].

*Image restoration results*

With knowledge of the PSF parameters eq. (6) $H_1$ in eq. (2) is fully specified and eq. (3) can be solved for the Raman map data of fig. 4a for different choices of $H_2$. The first choice was identity, resulting in a pixel size of $20nm \times 20nm$ ('20nm' for brevity, fig. 5a), identical to that in the observed image fig. 4a. The second choice was such that the resulting pixel size was $10nm \times 10nm$ ('10nm' for brevity, fig. 5b).

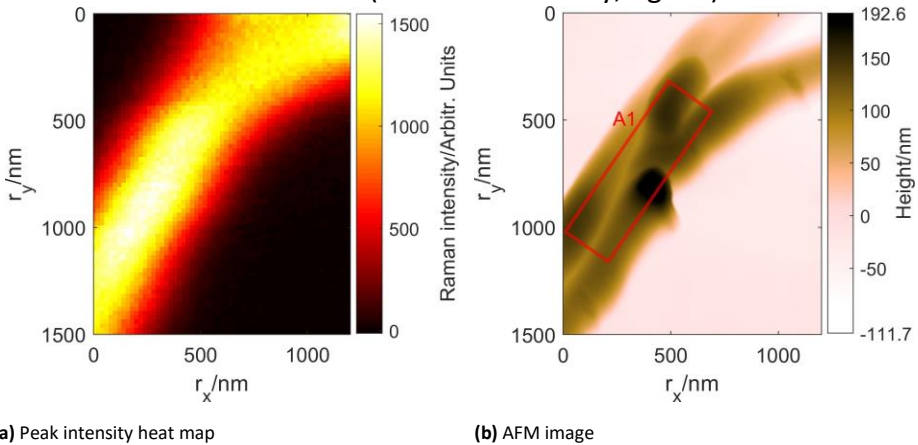

**(a)** Peak intensity heat map  **(b)** AFM image

**Figure 4.** (a) The Raman intensity of the spectral channel centred at the peak position of the dominant C=C peak ($1437 cm^{-1}$) plotted as a heat map over the sampling area. Pixel size is $20nm \times 20nm$ (b) An image of the sampling area obtained by atomic force microscopy. 2-3 PEDOT nanowires can be distinguished within the bundle. The red area 'A1' is marked for comparison with the SRIR images shown in fig. 5. The apparent width of the wires is broadened relative to their true width due to tip geometry effects[24,26]. Height, position, orientation, and alignment are imaged faithfully. Pixel size is $6nm \times 6nm$.

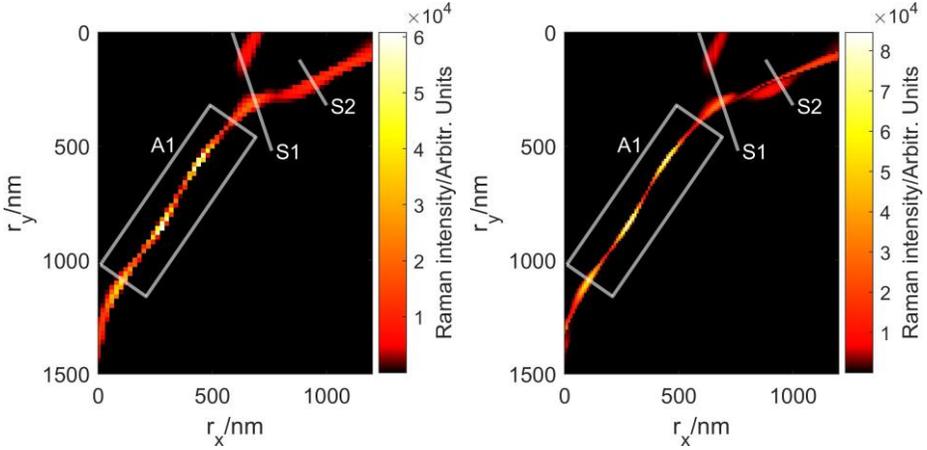

(a) SRIR with $20nm$ pixel size

(b) SRIR with $10nm$ pixel size

**Figure 5.** Super-resolution of the Raman heat map shown in fig. 4a for a resulting pixel sizes of (a) $20nm$ and (b) $10nm$. Two separate wires can be resolved at the top of each image. Line profiles along the sections S1 and S2 for both figures are shown in fig. 6.

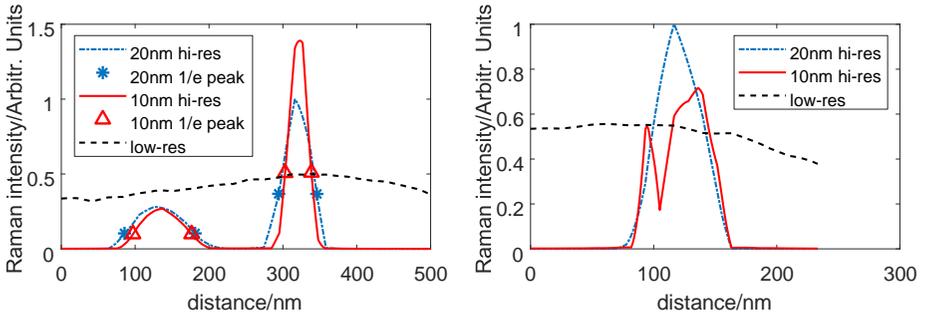

(a) section S1

(b) section S2

**Figure 6.** Line profiles along image sections of fig. 5: (a) Line intensity profile along section S1: Two objects are resolved, each of them appearing as a peak. The two peak maxima are separated by $189.8nm$ ($20nm$) and $186.0nm$ ($10nm$). The edges of the objects are located at the points where the intensity drops to $1/e$ times the peak intensity and are marked by stars ($20nm$) and triangles ($10nm$). The distance between the edges of the two objects are $113.2nm$ and $126.6nm$, respectively. (b) Line intensity profile along section S2: A single peak in the $20nm$ pixel size SRIR image can be resolved as two peaks in the $10nm$ pixel size SRIR image. The black dashed lines are the corresponding line profiles in the low-resolution image fig. 4a. The line profiles are obtained through linear interpolation of the respective images.

The properties of the SRIR results can be examined along the sections S1 and S2 shown in fig. 5. The corresponding line profiles obtained by linear interpolation are shown in fig. 6.

The super-resolved Raman images in fig. 5 do not match perfectly with the AFM image in fig. 4b but they illustrate the position, alignment, and orientation of the nanowires in good agreement with this image. They are much sharper and show more features than the observed low-resolution Raman image fig. 4a.

**Discussion**

*Sharpness and Resolution*

Downsampling in the forward operator of single-frame super-resolution renders the unconstrained optimisation problem eq. (3a) ill-posed. In multi-frame super-resolution this is compensated by the acquisition of multiple low-resolution images resulting in a total number of observed pixels that is equal to or larger than the number of pixels in the superresolved image, thus reinstating well-posedness. In the present work lost high spatial frequency information was recovered by the positivity constraint eq. (3b) and its implementation through primal-dual IPLS. This allowed for the reconstruction of the sharp high-resolution image fig. 5b with $10nm$ pixel size from just a single observation fig. 4a with $20nm$ pixel size. Comparison of fig. 5a and fig. 5b, as well as the line profiles of fig. 6, show that the super-resolved image with $10nm$ pixel size is much sharper than that with $20nm$ pixel size. While this is a very exciting result from an image restoration point of view, detailed analysis reveals the gain in sharpness does not necessarily corresponds to a gain in resolution.

The edges in the line profiles shown in figs. 6a and 6b have widths well below $50nm$ for both high-resolution images shown in fig. 5. However, this does not mean that the obtained resolvable distance is $50nm$ or smaller in each case. This is illustrated along the area A1 in fig. 5 where the SRIR results only show one nanowire with high intensity, while the corresponding AFM image in fig. 4b clearly shows 2-3 wires in the same area. The SRIR results show two separate nanowires only in areas where the two wires are separated further than the SRIR minimum resolvable distance, which can be estimated from the line profiles along the section S1 in fig. 6a. The two peaks observed correspond to two different nanowires that can be distinguished as two different objects. The peak to peak distance between their maximum intensities is $189.8nm$ and $186.0nm$ for $20nm$ and $10nm$ pixel size, respectively. For point-like objects this would give a measure of the spatial resolution. However, as known from the fabrication process and confirmed by AFM in fig. 4b we are dealing with objects of finite size and a more suitable measure would be their edge separation. Since their edges are not sharp, the edge positions were chosen to be at $1/e \approx 0.368$ times the peak intensities. The resulting edge separations are $113.2nm$ and $126.6nm$ for $20nm$ and $10nm$ pixel size, respectively, which is much larger than the observed width of edges for both cases. Although the width of edges and other features, which determines the sharpness of an image, puts a lower bound on the minimum resolvable distance, a measure for resolution, sharpness and resolution are two distinct properties and thus should be treated as such. Further, the apparent increase in sharpness in the $10nm$ pixel size super-resolved image compared to the $20nm$ pixel size one is not reflected in the achieved minimum resolvable distance.

This, however, should not obscure the satisfying result that the larger minimum resolvable distance, $126.6nm$, is about a factor 3 smaller than $FWHM_y = 362.9nm$, the smallest value obtained for the instrument's minimum resolvable distance.

*Comparison to AFM image*

Despite the enhanced resolution SRIR images still mismatch with the AFM image for a number of reasons. The minimum resolvable distance in AFM is well below $10nm$ (for objects of similar height), i.e. , much lower than the one achieved by SRIR. And even though AFM has much higher intrinsic resolution than optical microscopy, it cannot serve as a ground truth as AFM images can be blurred by tip geometry effects[24–26]. Furthermore, Raman spectroscopy and AFM measure fundamentally different material properties, so even if both techniques had indefinite resolution, identical images would not be expected (AFM, for instance, would not reveal any embedded nanoheterogeneity).

*Resolution enhancement and regularisation*

Resolution enhancement in SRIR is often achieved with the help of regularisation, which is simply adding a term of the form $\lambda\phi(x)$ to the optimisation criterion eq. (3a). Initially used as a heuristic to suppress noise artefacts in the restoration process, Bayesian analysis has managed to connect the regularisation function $\phi(\cdot)$ to prior knowledge about the super-resolved image[1]. The regularisation parameter $\lambda$ is usually chosen depending on the noise level in the observed low-resolution image. It has been recognised early that a large amount of prior information can be used to increase the resolution by a large amount[36]. An excellent example from fluorescence imaging demonstrates the power of aptly chosen regularisation while simultaneously hinting at the difficulties involved in arriving at such an apt choice[37]. The drawback of using regularisation is that prior information may not always be available to a sufficient degree, and even if it is, the choice of corresponding regularisation function $\phi(\cdot)$ and parameter $\lambda$ may not be straightforward.

The results of this paper were obtained without regularisation. Noise suppression and high spatial frequency estimation were achieved by the positivity constraint eq. (3b) and primal-dual IPLS[18] alone. This allows for SRIR of images for which no information about the super-resolved image is available a priori, apart from the generic assumption that the resulting pixel intensities should be positive. Simultaneously, if any additional prior knowledge about the super-resolved image is available, it can be straightforwardly implemented via a regularisation term in eq. (3a). This makes constrained optimisation via primal-dual IPLS a powerful and flexible tool for SRIR.

*Below the sampling step size*

The information received from a scanning stage microscope is band limited with the cut off frequency given by the inverse of the sampling step size. This means that by utilising the information received from the imaging process only it is impossible to resolve objects separated by less than a step size. By choosing appropriate constraints and regularisation, we are no longer restricted by the information theoretic limit as these provide a means to estimate Fourier coefficients above the cut off frequency which, when aptly chosen, can push resolution enhancement even further.

In the case of the present experiment this can be observed in the line profile along section S2, fig. 6b. The profile of the $20nm$ pixel size super-resolved image shows one peak whereas two peaks are clearly distinguishable for $10nm$ pixel size. It could be argued that the observed gap of about $20nm$ between the two peaks constitutes the minimum resolvable distance achieved by SRIR, which would indeed be an extraordinary and flattering result. But there is no experimental evidence to support such a claim and dependent on whether such a gap is desired or not it is either a feature or an artefact of the method employed.

The primary aim of this work was only to go beyond the intrinsic diffraction limit of the confocal Raman microscope. The occurrence of sub-step size separations in the super-resolved image fig. 5b should only give an idea of the possibilities when the full power of SRIR is unleashed.

**Conclusions**

We have successfully applied SRIR to Raman data of PEDOT nanowires. A conservative estimate suggests an enhancement in spatial resolution of about a factor 3 relative to the confocal Raman microscope's intrinsic resolution. The minimum resolvable distances of $360nm$ and higher in the raw data have been reduced to $125nm$ and lower by application of SRIR.

The gain in resolution through SRIR constitutes true super-resolution in the microscopist's sense of resolution as it has been demonstrated on objects that are unresolved in the raw data. The high-resolution images obtained through the proposed SRIR method, primal-dual IPLS, show satisfactory agreement when compared to AFM data of the same scene.

The raw data was acquired using a conventional Raman microscope equipped with a motorised piezo stage. No near-field spectroscopic techniques such as NFRS or TERS were involved, thus adding no complexity to the experimental setup and avoiding any kind of electronic interaction with the sample material.

In addition we demonstrated and discussed the potential of primal-dual IPLS to resolve distances below the raw data sampling step size of $20nm$. Faithful SRIR at such small scales would require thorough thermal and mechanical stabilisation during the measurement process and will be the subject of a future investigation.

Further, the SRIR results of this paper were generated by processing data from only one spectral channel of the Raman microscope. Data from multiple channels would allow for the retrieval of chemical information with high spatial resolution and perhaps for further reduction of the minimum resolvable distance through SRIR with primal-dual IPLS. These would be desirable results as they would not require any additional measurements or experimental complications and will be addressed in future publications.

## Declaration of conflicting interests

There are no conflicts to declare.

## Funding

This work has received funding from the European Union's Horizon 2020 research and innovation programme under the Marie Skłodowska-Curie grant agreement No 642742.

## Supplemental material

### The point-spread function and resolution

The point-spread function (PSF) of an optical microscope is the image observed from a single point-like scatterer. Its form and width determines the instrument's resolution. The minimum resolvable distance $\Delta r$ in a confocal microscope is given by the diffraction or Rayleigh limit[35,38]

$$\Delta r = 0.44 \frac{\lambda}{n\,NA}, \tag{8}$$

where $\lambda$ is the wavelength of the observed light, $n$ the refractive index of the medium and *NA* the numerical aperture of the objective lens. For a system with $\lambda = 532nm$, $n = 1$ and $NA = 0.85$ eq. (8) yields $\Delta r = 275nm$.

This constitutes the theoretically optimal case for which the PSF can be found in[35,39]. The shape of the PSF is tightly connected to the resolution of an optical instrument. Airy described the diffraction pattern of a self-luminous point-source (star) under an object-glass with circular aperture[40], which is just the PSF of his instrument. Rayleigh later proposed to define the minimum resolvable distance $\Delta r$ as the lateral distance from the optical axis for which the Airy diffraction pattern attains its first minimum[38]. The intensity dip between two self-luminous point-like objects of equal intensity (wide-field illumination) separated by $\Delta r$ is about 26.4% the maximum intensity[35,41]. Rayleigh's original definition of $\Delta r$ is specific to the Airy diffraction pattern, but it has been extended to instruments that do not exhibit an Airy diffraction pattern as the lateral separation of two equiluminous point sources that yields a 26.4% dip between the two maximum intensities observed[35,41]. Modern descriptions of resolution, however, tend away from absolute values for $\Delta r$ and rather state it for a given value of contrast[35,41].

For a Gaussian PSF like in eq. (4) Rayleigh's $\Delta r$ along each axis can be approximated by

$$\Delta r_{\{x,y\}} \approx 2.80\sigma_{\{x,y\}}. \qquad (9)$$

Another commonly used experimental measure is the full width at half maximum (FWHM) of the PSF. The FWHM for eq. (4) is given by

$$FWHM_{\{x,y\}} = 2\sqrt{2\log(2)}\sigma_{\{x,y\}} \qquad (10a)$$
$$\approx 2.35\sigma_{\{x,y\}}. \qquad (10b)$$

Two equiluminous sources separated by FWHM will show a dip of 7.3% of the maximum intensity when observed through a microscope with Gaussian PSF. A separation of less than $2\sigma_{\{x,y\}}$ for the PSF eq. (4) produces no dip between two objects, regardless of their relative luminosity, and is thus unresolvable.

## Super-resolution of simulated test images

The primal-dual algorithm used in this paper[18] for solving the interior-point least squares (IPLS) problem eq. (3) can be tested on computer generated images. From a given ground truth an observed image can be calculated by application of the observation model eq. (1). The observed image can then be super-resolved and compared to the given ground truth. Two simple cases are considered:

i) two identical parallelograms aligned parallel to each other with shorter edges of $80nm$, longer edges of $870nm$, and minimum separation of $110nm$, fig. 7a; ii) one parallelogram with shorter edge of $270nm$ and longer edge of $870nm$, fig. 8a. Case (ii) is the same as case (i) but with the gap between the two parallelograms filled. Both generated ground truths have a pixel size of $10nm \times 10nm$.

Each ground truth was convolved with a Gaussian PSF given in eq. (4) with $\sigma_x = 200nm$ ($FWHM_x = 471nm$) and $\sigma_y = 150nm$ ($FWHM_y = 353nm$), downsampled to images of pixel size $20nm \times 20nm$, and added to i.i.d. Gaussian noise with a signal-to-noise ratio (SNR) of $30dB$ to yield the respective observed images, figs. 7b and 8b. Super-resolution image restoration (SRIR) with IPLS was then performed with two different forward operators on each of the simulated observations. The first choice of forward operator was convolution with the PSF used in generating the simulated observations, resulting in high-resolution images of pixel size $20nm \times 20nm$, figs. 7c and 8c. The second choice of forward operator was the full forward operator used for generating the simulated observations, i.e. convolution with the PSF and downsampling by a factor of 2 along each image axis, resulting in high-resolution images of pixel size $10nm \times 10nm$, figs. 7d and 8d.

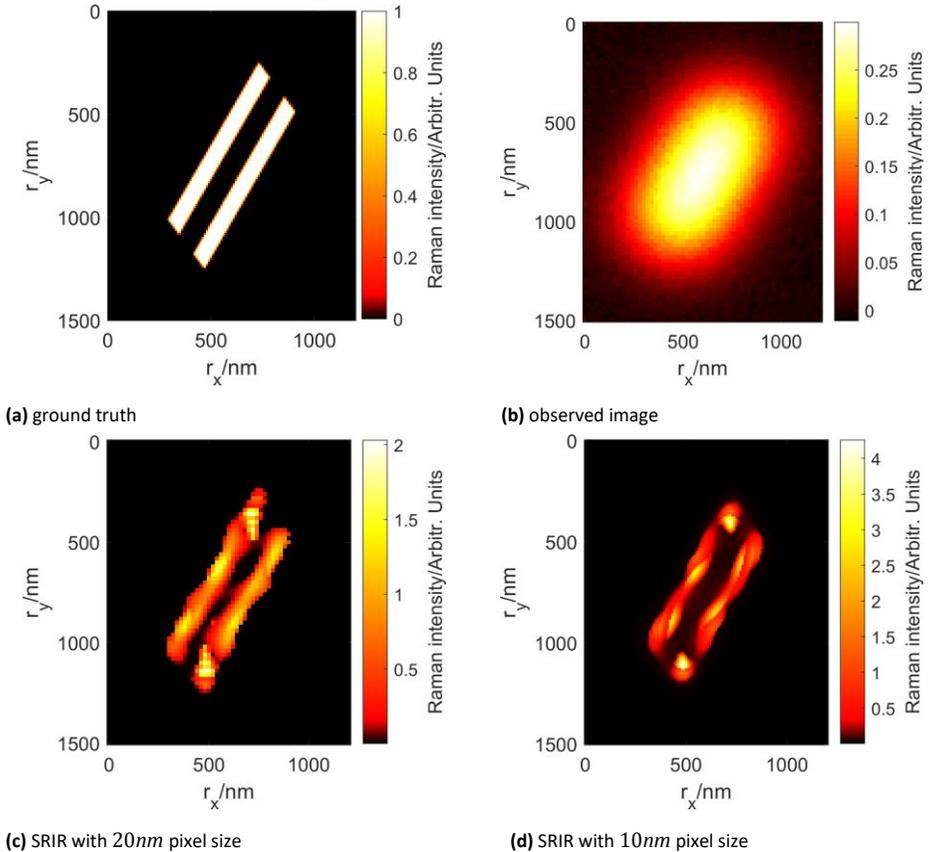

**(a)** ground truth  **(b)** observed image

**(c)** SRIR with $20nm$ pixel size  **(d)** SRIR with $10nm$ pixel size

**Figure 7.** Demonstration of IPLS on a numerical test image: (a) Simulated ground truth with pixel size of $10nm \times 10nm$; (b) observed image after convolution with the PSF of eq. (4) with $\sigma_x = 200nm$ and $\sigma_y = 150nm$, downsampling (pixel size $20nm \times 20nm$), and the addition of Gaussian white noise with SNR = $30dB$; (c) result of IPLS without up-/downsampling (pixel size $20nm \times 20nm$); (d) result of IPLS with up-/downsampling (pixel size $10nm \times 10nm$)

*Winterauer et al.* 21

The results of IPLS, figs. 7c and 7d and figs. 8c and 8d, are by no means perfect restorations, as can be seen by comparison with the respective ground truths fig. 7a and fig. 8a. Because the size of the objects in the ground truth images is smaller than the width of the PSF along at least one dimension a perfect restoration would not be expected. However, the resulting super-resolved images figs. 7c and 7d and figs. 8c and 8d are qualitatively very different, in resemblance of the respective ground truth, despite the large similarity between the observed images fig. 7b and fig. 8b.

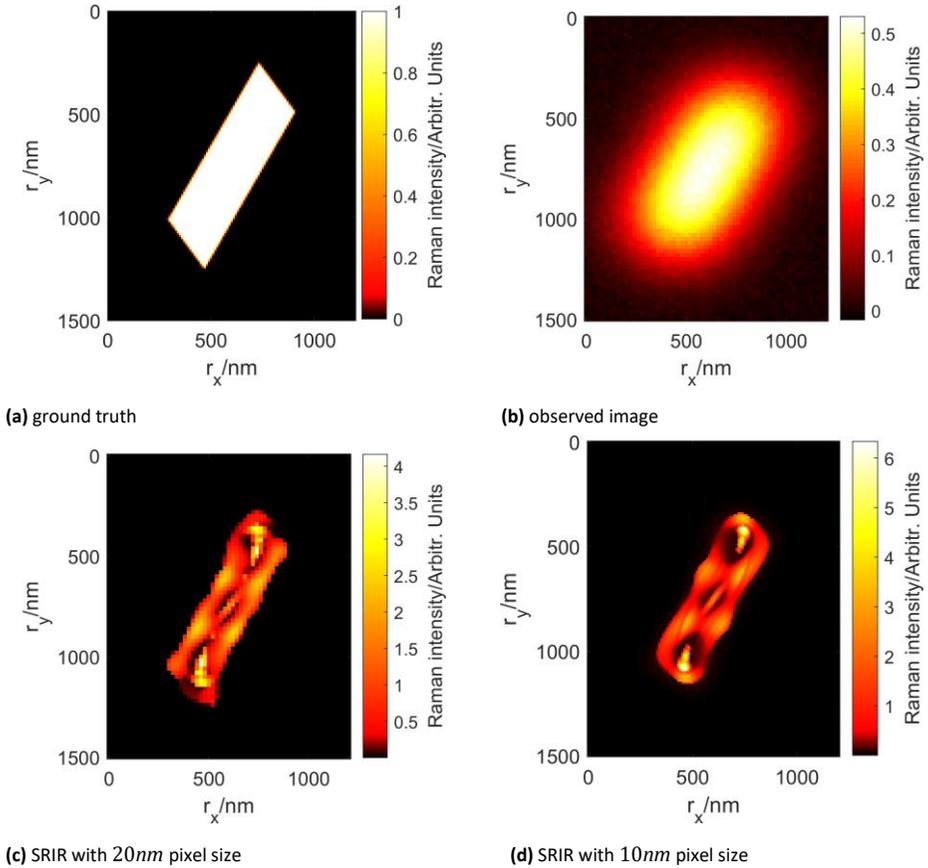

**(a)** ground truth

**(b)** observed image

**(c)** SRIR with $20nm$ pixel size

**(d)** SRIR with $10nm$ pixel size

**Figure 8.** Demonstration of IPLS on a numerical test image: (a) Simulated ground truth with pixel size of $10nm \times 10nm$; (b) observed image after convolution with the PSF of eq. (4) with $\sigma_x = 200nm$ and $\sigma_y = 150nm$, downsampling (pixel size $20nm \times 20nm$), and the addition of Gaussian white noise with SNR $= 30dB$; (c) result of IPLS without up-/downsampling (pixel size $20nm \times 20nm$); (d) result of IPLS with up-/downsampling (pixel size $10nm \times 10nm$)